\documentclass[aps,prd,amsmath,amssymb,preprintnumbers,nofootinbib,a4paper,11pt]{revtex4}
\pdfoutput=1
\usepackage{amsthm}
\usepackage{graphicx}
	\graphicspath{{./NewFig/}}
\usepackage{url}
\usepackage[bookmarks, pagebackref=false]{hyperref}
\usepackage{color}
	\definecolor{rossoCP3}{cmyk}{0,.88,.77,.40}
		\definecolor{graa}{rgb}{0.8,0.8,0.8}
		\definecolor{blaa}{rgb}{0.2,0.2,0.6}
		\hypersetup{
			colorlinks, 
			bookmarksopen, 
			bookmarksnumbered,
			citecolor=blaa, 		%color of links to bibliography
			linkcolor=rossoCP3,	%color of internal links
			urlcolor=rossoCP3,			%color of external links 
			}
\usepackage{dcolumn}% Align table columns on decimal point
\usepackage{bm}% bold math
\usepackage{bbm}
\usepackage{pxfonts}

\usepackage{epsfig}
\usepackage{placeins}

\usepackage{bbold}

\usepackage[ margin=5pt, font=small,labelfont=bf, justification=raggedright]{caption}
\usepackage{youngtab}
\usepackage{slashed}
\usepackage{subfig}

\newcommand{\beq}{\begin{eqnarray}}
\newcommand{\eeq}{\end{eqnarray}}

\newcommand{\bmp}{\noindent\begin{minipage}{16cm}}
\newcommand{\emp}{\end{minipage}\vskip 7mm} % 7mm untightened

\def\lsim{\mathrel{\rlap{\lower4pt\hbox{\hskip1pt$\sim$}}
    \raise1pt\hbox{$<$}}}                % less than or approx. symbol
\def\gsim{\mathrel{\rlap{\lower4pt\hbox{\hskip1pt$\sim$}}
    \raise1pt\hbox{$>$}}}                % greater than or approx. symbol

\begin{document}
%%%%%%%%%%%%%%%%%%%%%%%%%%%%%%%%%%%%%%%%%%%%%%%%%%%%%%%%%%%%%%%%%%%%%%%%%%%

\title{\Large  Consistent Perturbative Fixed Point Calculations in QCD and SQCD}
 \author{Thomas A. Ryttov}\email{ryttov@cp3.dias.sdu.dk} 
  \affiliation{
{ \color{rossoCP3}  \rm CP}$^{\color{rossoCP3} \bf 3}${\color{rossoCP3}\rm-Origins} \& the Danish Institute for Advanced Study {\color{rossoCP3} \rm Danish IAS},\\ 
University of Southern Denmark, Campusvej 55, DK-5230 Odense M, Denmark.
}

%%%%%%%%%%%%%%%%%%%%%%%%%%%%%%%%%%%%%%%%%%%%%%%%%%%%%%%%%%%%%%%%%%%%%%%%%%%%%%%%%%%%%%%%
\begin{abstract}

We suggest how to consistently calculate the anomalous dimension $\gamma_*$ of the $\bar{\psi}\psi$ operator in finite order perturbation theory at an infrared fixed point for asymptotically free theories. If the $n+1$ loop beta function and $n$ loop anomalous dimension are known then $\gamma_*$ can be calculated exactly and fully scheme independently through $O(\Delta_f^n )$ where $\Delta_f = \bar{N_f} - N_f$ and $N_f$ is the number of flavors and $\bar{N}_f$ is the number of flavors above which asymptotic freedom is lost. For a supersymmetric theory the calculation preserves supersymmetry order by order in $\Delta_f$.  We then compute $\gamma_*$ through $O(\Delta_f^2)$ for supersymmetric QCD in the $\overline{\text{DR}}$ scheme and find that it matches the exact known result. We find that $\gamma_*$ is astonishingly well described in perturbation theory already at the few loops level throughout the entire conformal window. We finally compute $\gamma_*$ through $O(\Delta_f^3)$ for QCD and a variety of other non-supersymmetric fermionic gauge theories. Small values of $\gamma_*$ are observed for a large range of flavors. 

 \vskip .1cm
{\footnotesize  \it Preprint:  CP$^3$-Origins-2016-016  DNRF 90\ \& DIAS-2016-16}
 \end{abstract}

\maketitle

\newpage
     
\section{Introduction}

Exact and scheme independent results for strongly interacting systems are hard to come by. Although some physical quantity should in principle be scheme independent one is usually limited by finite order perturbation theory in the practical calculation and therefore scheme dependence of the final result is inevitably induced. Quantum Chromo Dynamics (QCD) at an interacting fixed point provides a perfect framework in which we shall address this concern.

Of our interest is the calculation of the anomalous dimension of the bilinear operator $\bar{\psi}\psi$ of some gauge theory with a single set of massless fermion representations $\psi_f, \ f=1,\ldots,N_f $ at an interacting fixed point. If the gauge group is $SU(3)$  and the fermion representation is the fundamental then the theory is QCD with $N_f$ flavors. To see that the anomalous dimension at a fixed point is a physical scheme independent quantity we first recall that the anomalous dimension and the beta function are defined via
\begin{eqnarray}
\gamma = - \frac{d \ln Z_{\bar{\psi}\psi}}{d \ln\mu} \ , \qquad \beta(\alpha) =\frac{d \alpha}{d \ln \mu}
\end{eqnarray}
where $Z_{\bar{\psi}\psi}$ is the renormalization constant of the bilinear operator $\bar{\psi}\psi$ and $\alpha = \frac{g^2}{4\pi}$ with $g$ being the gauge coupling. Assume now that we change the scheme in which we have calculated these renormalization group functions to another alternative scheme. This corresponds to a change of variables with the alternative scheme having a different definition of the gauge coupling $\tilde{\alpha}(\alpha)$. It follows that the anomalous dimension and the beta function take the simple forms
\begin{eqnarray}\label{eq:schemes}
\tilde{\beta}(\tilde{\alpha}) = \frac{\partial \tilde{\alpha}}{\partial \alpha} \beta(\alpha)   \ , \qquad  \tilde{\gamma}(\tilde{\alpha}) = \gamma(\alpha) + \beta(\alpha) \frac{\partial \ln F}{\partial \alpha}
\end{eqnarray}
in the alternative scheme where $F=Z_{\bar{\psi}\psi} \tilde{Z}_{\bar{\psi}\psi}^{-1}$. Therefore the existence of a fixed point is scheme independent, i.e. if $\beta(\alpha_*)=0$ for some $\alpha_*$ then also $\tilde{\beta}(\tilde{\alpha}_*)=0$ for some $\tilde{\alpha}_*$, and the value of the anomalous dimension at this fixed point is scheme independent $\tilde{\gamma}(\tilde{\alpha}_*) = \gamma(\alpha_*) \equiv \gamma_*$. 

Unfortunately in most cases we are limited by only knowing the beta function and anomalous dimension to finite order in perturbation theory. For  gauge theories with a single fermion representation they are known to four loop order in the $\overline{\text{MS}}$ scheme  \cite{vanRitbergen:1997va} while for their supersymmetric cousin theories they are known to three loop order in the $\overline{\text{DR}}$ scheme \cite{Jack:1996vg}. This limitation is the source of an induced problem we wish to solve.

Suppose that we expand the beta function and anomalous dimension in the gauge coupling as
\begin{eqnarray}
\frac{2\pi}{\alpha^2} \beta (\alpha) &=& - \sum_{i=0}^{\infty} \beta_{i} \left( \frac{\alpha}{2\pi} \right)^{i} \\
\gamma(\alpha) &=& \sum_{i=1}^{\infty} \gamma_i \left( \frac{\alpha}{2\pi}  \right)^{i}
\end{eqnarray}

We shall take the trivial fixed point to be in the ultraviolet such that the theory is asymptotically free. This is achieved provided $\beta_0  >0 $, i.e. for a sufficiently small number of flavors \cite{Gross:1973id}. For a number of flavors just below the critical value where $\beta_0$ turns positive the second beta function coefficient $\beta_1<0$ \cite{Caswell:1974gg}. Suppose that we only have available the first two coefficients of the beta function and anomalous dimension. Then the theory has a non-trivial positive fixed point located at
\begin{eqnarray}
\frac{\alpha_*}{2\pi} &=& -  \frac{\beta_0}{\beta_1} 
\end{eqnarray}
Evaluating the anomalous dimension at this fixed point yields
\begin{eqnarray}
\gamma_* &=& - \frac{\gamma_1 \beta_0}{\beta_1}   + \frac{\gamma_2\beta_0^2}{\beta_1^2}  
\end{eqnarray}
Now since $\beta_0$, $\beta_1$ and $\gamma_1$ are all scheme independent whereas $\gamma_2$ is scheme dependent also $\gamma_*$ must be scheme dependent. This seems to be in contradiction with our general observation above. The scheme dependent contamination of $\gamma_*$ is of course due to our artificial truncation of the beta function and anomalous dimension. Also it should be clear that this problem persists to any finite order in perturbation theory. 

The value of $\gamma_*$ obtained in his way at three and four loops in the $\overline{\text{MS}}$ scheme has been studied in \cite{Ryttov:2010iz} and in various other schemes in \cite{Gracey:2015uaa}. However no method for disentangling the induced scheme dependent error from the perturbative approximation was given. We wish to outline a procedure for how to obtain the scheme independent information which still should be hidden in $\gamma_*$ computed using finite order perturbation theory. 

Finally we note that if the theory is supersymmetric then the above calculation of $\gamma_*$ also generically breaks supersymmetry which is again due to the artificial truncation of the perturbative expansion. This second induced problem we similarly intend to solve. In \cite{Ryttov:2012qu} the two and three loop evaluation of $\gamma_*$ was studied with the unsatisfactory conclusion that $\gamma_*$ turned negative at three loops for a number of flavors not very much below the critical value where asymptotic freedom is lost. As we will see below the culprit can now be identified to be induced scheme dependence and explicit supersymmetry breaking.

\section{Consistently Computing The Anomalous Dimension}

We suggest instead to calculate $\gamma_*$ as a series expansion in  $\Delta_f \equiv \bar{N}_f - N_f$
\begin{eqnarray}
\gamma_* &=& \sum_{i=1}^{\infty} c_i \Delta_f^i
\end{eqnarray} 
where $\bar{N}_f$ is the fixed number of flavors above which asymptotic freedom is lost. For illustrative purposes we shall view the number of flavors as a continuous parameter although ultimately only integer values are of physical interest.

Such an expansion has several attractive features. Of course if we know the beta function and anomalous dimension to all orders then we can in principle exactly compute each coefficient $c_i$. Each coefficient must then also be scheme independent: Imagine that we have computed $\tilde{\gamma}_*$ in another scheme then we can similarly expand it in $\Delta_f$ (which is scheme independent) but with coefficients $\tilde{c}_i$. Forming the difference $\gamma_* - \tilde{\gamma}_*=0$ we find that this can only be satisfied provided $c_i-\tilde{c}_i =0$, i.e. the coefficients $c_i$ are scheme independent. 

Fortunately as we shall see each $c_i$ will not depend on the full beta function and anomalous dimension. In fact $c_i$ will only depend on information stored in the $i+1$ loop beta function and $i$ loop anomalous dimension. It will not receive contributions from higher loop orders. Hence if it happens that we only know the beta function through $n+1$ loop order and the anomalous dimension through $n$ loop order (we will refer to this as an $(n+1,n)$ loop order computation) this will still enable us to compute $\gamma_*$ in an \emph{exact} and \emph{scheme independent} manner through $O(\Delta_f^n)$. This we will now show by providing a method for how to in principle compute the coefficients $c_i$ to any desired order.

As we go to higher loop orders finding fixed points of the beta function becomes increasingly difficult. In addition new non-physical solutions appear which we will have to discard. We are only interested in the specific fixed point solution which is small in the limit where $\Delta_f$ is small. Hence we will make the following ansatz for the fixed point solution and expand it as follows
\begin{eqnarray}
\frac{\alpha_*}{2\pi} &=& \sum_{i=1}^{\infty}  a_i \Delta_f^i
\end{eqnarray}
Our first job is  to find the coefficients $a_i$. Evaluating the beta function at this fixed point gives
\begin{eqnarray}\label{zero}
0&=& \sum_{i=0}^{\infty} \beta_i \left( \sum_{j=1}^{\infty} a_j \Delta_f^j \right)^i  = k_1 \Delta_f + k_2 \Delta_f^2 + k_3 \Delta_f^3 + O(\Delta_f^4)
\end{eqnarray}
where we have expanded in $\Delta_f$. The first constant term in the expansion vanishes since this is just $\beta_0$ evaluated at $N_f=\bar{N}_f$. The first three coefficients are
\begin{eqnarray}
k_1 &=& a_1 \bar{\beta}^{(0)}_{1} - \bar{\beta}^{(1)}_{0} \\
k_2 &=& a_2 \bar{\beta}^{(0)}_{1} + a_1^2 \bar{\beta}^{(0)}_{2} - a_1 \bar{\beta}^{(1)}_{1}  \\
k_3 &=& a_3 \bar{\beta}^{(0)}_{1} +2a_1a_2 \bar{\beta}^{(0)}_{2}  -a_2 \bar{\beta}^{(1)}_{1} - a_1^2 \bar{\beta}^{(1)}_{2} +a_1^3 \bar{\beta}^{(0)}_{3}
\end{eqnarray}
and we have defined 
\begin{eqnarray}
\bar{\beta}_i^{(n)} &=& \frac{\partial^n  \beta_i}{\partial N_f^n}_{|N_f=\bar{N}_f}
\end{eqnarray}
We have also used the fact that $\bar{\beta}^{(2)}_{0} = \bar{\beta}^{(3)}_{0} = \bar{\beta}^{(2)}_{1} =0 $ since the first two coefficients only contain terms proportional to $N^0_f$ or $N_f^1$. 

Since we take $\Delta_f$ to be an arbitrary positive number Eq. \ref{zero} can only be satisfied provided each coefficient $k_i=0$. These conditions will give us a set of equations that can be used to solve for the coefficients $a_i$. Setting $k_1=0$ we can solve for $a_1$. Setting $k_2=0$ we can solve $a_2$ since we know $a_1$ while setting $k_3=0$ we can solve for $a_3$ since we know $a_1$ and $a_2$. This gives
\begin{eqnarray}\label{eq:a1a2}
a_1 &=& \frac{\bar{\beta}^{(1)}_{0}}{\bar{\beta}^{(0)}_{1}} \\
 a_2 &=& \left( \bar{\beta}^{(0)}_{1} \bar{\beta}^{(1)}_{1} - \bar{\beta}^{(1)}_{0} \bar{\beta}^{(0)}_{2} \right) \frac{\bar{\beta}^{(1)}_{0}}{ \left.  \bar{\beta}^{(0)}_{1}  \right.^3 } \\
a_3 &=& \left( \left. \bar{\beta}^{(0)}_{1} \right.^2 \left. \bar{\beta}^{(1)}_{1} \right.^2 - 3 \bar{\beta}^{(1)}_{0} \bar{\beta}^{(0)}_{1} \bar{\beta}^{(1)}_{1}   \bar{\beta}^{(0)}_{2} +2 \left. \bar{\beta}^{(1)}_{0}  \right.^2 \left. \bar{\beta}^{(0)}_{2}  \right.^2  \right.     \left. +  \bar{\beta}^{(1)}_{0} \left. \bar{\beta}^{(0)}_{1} \right.^2 \bar{\beta}^{(1)}_{2} - \bar{\beta}^{(0)}_{1} \left. \bar{\beta}^{(1)}_{0} \right.^2 \bar{\beta}^{(0)}_{3}   \right)  \frac{\bar{\beta}^{(1)}_{0}}{ \left. \bar{\beta}^{(0)}_{1}  \right.^5} \label{eq:a3}
\end{eqnarray}
The pattern for solving for the coefficients $a_i$ one at a time continues and in principle allows us to solve for the fixed point to any desired order in $\Delta_f$ in a quite straight forward manner. The coefficient $k_i$ is of the form
\begin{eqnarray}
k_i &=& a_i \bar{\beta}^{(0)}_{1} + f( a_{1}, \ldots,a_{i-1},\bar{\beta}_0,\ldots,\bar{\beta}_{i})
\end{eqnarray}
where $f$ is a combinatorial function that in principle can be computed using Fa\`a di Bruno's formula for calculating the $i$'th derivative of a composite function. Setting $k_i=0$ allows to solve for $a_i$ assuming that we have already solved for $a_1,\dots,a_{i-1}$.  For our purposes however it suffices to only know explicitly the first three coefficients $a_i,\ i=1,2,3$. 

We observe that $a_1$ only depends on the two loop coefficients, $a_2$ only depends on the three loop coefficients and $a_3$ only depends on the four loop coefficients. In general $a_i$ will only depend on the first $i+1$ loop coefficients. 

We now take the last step and evaluate the anomalous dimension at the fixed point
\begin{eqnarray}\label{eq:ano}
\gamma_* &=& \sum^{\infty}_{i=1} \gamma_i \left( \sum^{\infty}_{j=1} a_j \Delta_f^j  \right)^i = c_1 \Delta_f +c_2 \Delta^2_f + c_3 \Delta^3_f + O(\Delta_f^4)
\end{eqnarray}
where we have again expanded in $\Delta_f$. The first three coefficients are
\begin{eqnarray}
c_1 &=&  a_1 \bar{\gamma}^{(0)}_{1} \\
c_2  &=& a_2 \bar{\gamma}^{(0)}_{1} + a_1^2 \bar{\gamma}^{(0)}_{2} - a_1 \bar{\gamma}^{(1)}_{1}   \\
c_3 &=& a_3 \bar{\gamma}^{(0)}_{1}  + 2a_1a_2  \bar{\gamma}^{(0)}_{2} + a_1^3  \bar{\gamma}^{(0)}_{3} - a_1^2  \bar{\gamma}^{(1)}_{2}
\end{eqnarray}
with 
\begin{eqnarray}
\bar{\gamma}^{(n)}_{i} &=& \frac{\partial^n \gamma_i}{ \partial N_f^n}_{|N_f=\bar{N}_f}
\end{eqnarray} 
In writing these coefficients we have used the fact that $\gamma_1$ does not depend on the number of flavors. Now it is clear that $c_1$ will only depend on the $(2,1)$ loop coffificients via $a_1,\gamma_1$. Also $c_2$ will only depend on the $(3,2)$ loop coefficients via $a_1,a_2,\gamma_1,\gamma_2$ while $c_3$ will only depend on the $(4,3)$ loop coefficients via $a_1,a_2,a_3,\gamma_1,\gamma_2,\gamma_3$. In general $c_i$ will only depend on the $(i+1,i)$ loop coefficients of the beta function and anomalous dimension. They will not receive any further corrections from higher loops. Therefore if we have available the $n+1$ loop beta function and the $n$ loop anomalous dimension we can compute $\gamma_*$ through $O(\Delta_f^n)$ in an exact and fully scheme independent manner. This is precisely what we wanted.

\section{Supersymmetric QCD}

We are now at a point where we can put our explicit formula and results for $\gamma_*$ to work. Before discussing QCD we will take a small departure and test our investigations against exact known results in supersymmetric QCD. The same line of reasoning as above will lead us to Eq. \ref{eq:ano} but with $N_f$ now counting the number of superflavors and $\bar{N}_f=\frac{3}{2}\frac{C_A}{T_r}$ denoting the number of superflavors above which asymptotic freedom is lost. The group factors $C_r$ and $T_r$ are respectively the quadratic Casimir and trace normalization factor for the representation $r$  and $A$ denotes the adjoint representation. Using the $(3,2)$ loop coefficients calculated in the $\overline{\text{DR}}$ scheme in \cite{Jack:1996vg} the coefficients $c_1$ and $c_2$ can be found and for the value of $\gamma_*$ we therefore arrive at 
\begin{eqnarray}\label{DR}
\gamma_* &=& \frac{2T_r}{3C_A} \Delta_f + \left( \frac{2T_r}{3C_A} \right)^2 \Delta_f^2 + O(\Delta_f^3)
\end{eqnarray}
This result is directly comparable to the exact result (computed through a different scheme) already known to exist \cite{Seiberg:1994pq} and which we write in the following suggestive form
\begin{eqnarray}
\gamma_* &=& \frac{\frac{2T_r}{3C_A}\Delta_f}{1-\frac{2T_r}{3C_A}\Delta_f} 
\end{eqnarray}
Through $O(\Delta_f^2)$ we see there is complete agreement with the result obtained in the $\overline{\text{DR}}$ scheme, Eq. \ref{DR}, as there should be. 

\begin{figure}[t!]
  \centering
     \includegraphics[width=0.35\textwidth]{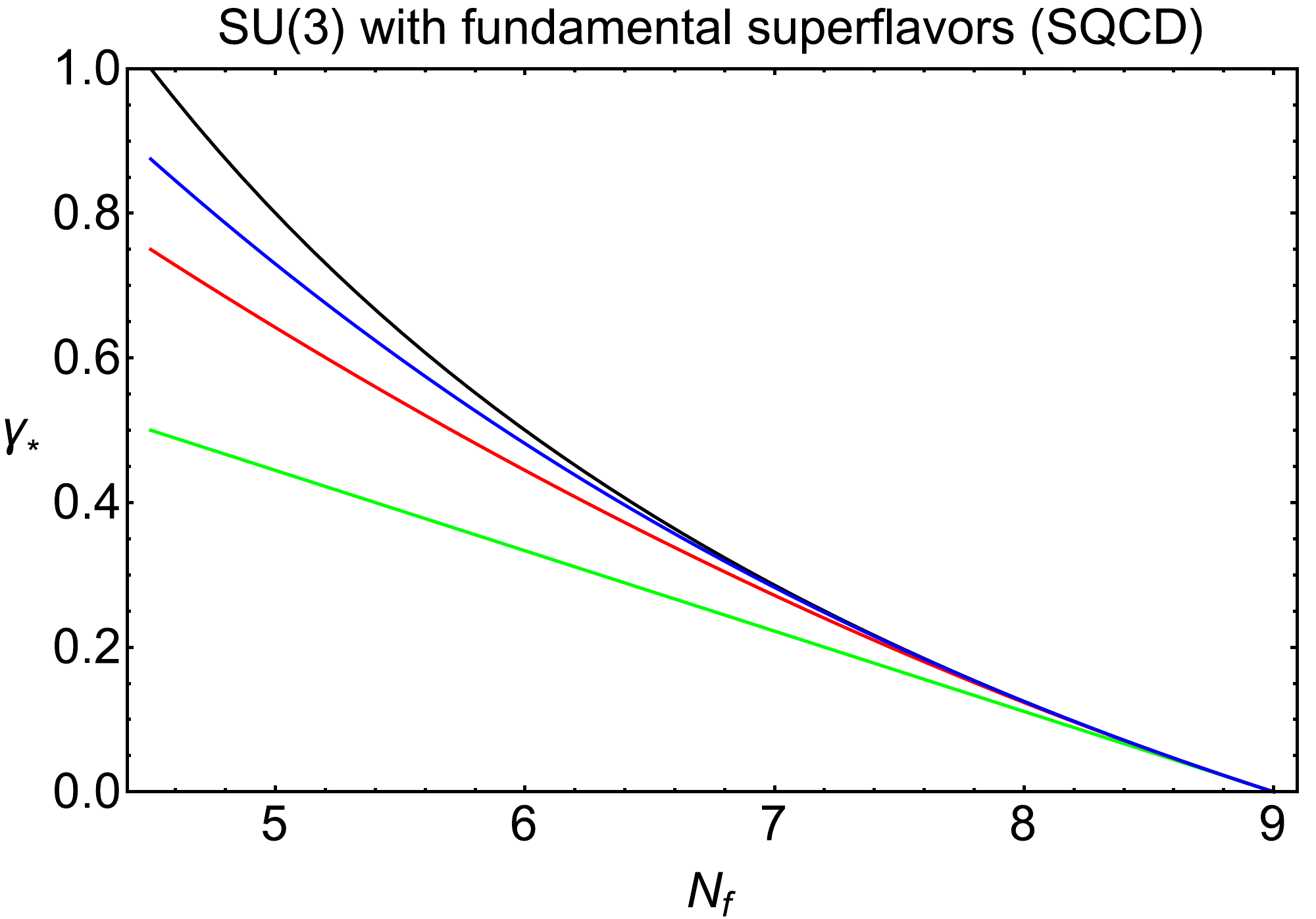}
     \qquad\qquad 
      \includegraphics[width=0.35\textwidth]{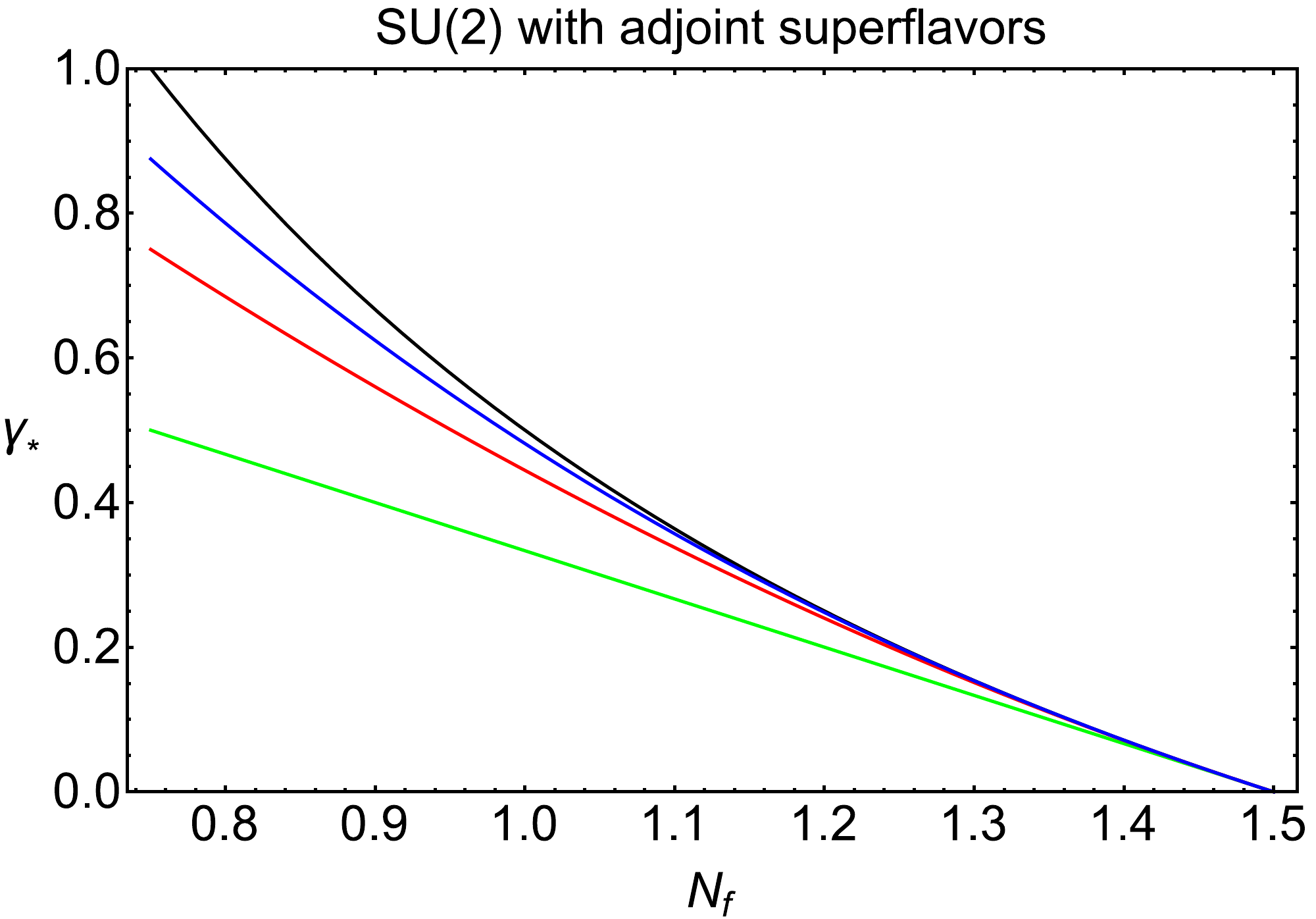}
   \caption{The value of $\gamma_*$ for supersymmetric QCD with gauge group $SU(3)$ and $N_f$ fundamental superfields (left panel) and gauge group $SU(2)$ and $N_f$ adjoint superfields (right panel). The green curve is $O(\Delta_f)$, the red curve is $O(\Delta_f^2)$, the blue curve is $O(\Delta_f^3)$ and the black curve is the exact result.}
   \label{SUSY}
\end{figure}

It is instructive to plot $\gamma_*$ as a function of the number of superflavors. This we do in Fig. \ref{SUSY} for an $SU(3)$ gauge theory with fundamental matter (left panel) and for an $SU(2)$ gauge theory with adjoint matter (right panel). In both plots $\gamma_*$ is plotted through $O(\Delta_f)$ (green), through $O(\Delta_f^2)$ (red) and exactly (black). The blue curve is $\gamma_*$ through $O(\Delta_f^3)$ obtained from the exact result and which we now know must correspond to a $(4,3)$ loop computation. In fact knowing the exact result our investigations allow us predict that an $(n+1,n)$ loop computation must yield
\begin{eqnarray}
c_n &=& \left( \frac{2T_r}{3C_A} \right)^n
\end{eqnarray}

It is clear that perturbation theory provides a remarkably accurate estimate already at $O(\Delta_f^3)$ (blue curve) as compared to the exact result (black curve). To our surprise the physics at the fixed point seems to be \emph{very well} described by higher order perturbation theory \emph{throughout the entire conformal window}, $0<\gamma_* <1$. 

Initially when we set out to perform the computation of $\gamma_*$ one might have feared about potential problems of convergence of the perturbative expansion since $\Delta_f$ is typically not a small number. However there is no need for such worries. In fact by inspecting the exact result we see that $\gamma_*$ has a series expansion in $\Delta_f$ provided $\frac{2T_r}{3C_A} \Delta_f <1$ which corresponds to $N_f>0$. Therefore the series expansion formally exists for all asymptotically free theories. Of course it only makes sense to calculate $\gamma_*$ for the theories that actually reach the fixed point and preserve unitarity \cite{Mack:1975je} $0< \gamma_*<1$ corresponding to $ \frac{3C_A}{4T_r} <N_f<\frac{3C_A}{2T_r}$ \cite{Seiberg:1994pq}.

At last we remark that although an artificial truncation of the beta function and anomalous dimension explicitly breaks supersymmetry it should be clear that our method for computing $\gamma_*$ has the strength that supersymmetry is preserved order by order.

Alternative approaches for comparing perturbative calculations to the exact result exist in the literature but they seem to either be scheme dependent, break supersymmetry or both \cite{Ryttov:2012qu}, \cite{Choi:2015azs}.

\section{QCD}

Having seen how accurately $\gamma_*$ is described by a $(4,3)$ loop computation in supersymmetric QCD we turn our attention to QCD for which the similar computation can be directly done. Here the critical number of flavors below which the theory is asymptotically free is $\bar{N}_f = \frac{11C_A}{4T_r}$. Using the $(4,3)$ loop coefficients of the beta function and anomalous dimension found in \cite{vanRitbergen:1997va} we can calculate $\gamma_*$ through $O(\Delta_f^3)$ as done above. Direct evaluation gives  
\begin{eqnarray}
c_1 &=& \frac{8T_r C_r}{C_A (7C_A + 11C_r)} \ , \qquad c_2 = \frac{4T_r^2C_r (35C_A^2 +636C_AC_r + 352 C_r^2) }{3C_A^2(7C_A+11C_r)^3} \\
c_3 &=& \frac{4T_r C_r}{81C_A^4 (7C_A + 11C_r)^5 } \left( -55419 T_r^2 C_A^5   + 432012T^2C_A^4 C_r  + 5632T_r^2  C_r \frac{d_A^{abcd}d_A^{abcd}}{d_A} (-5 +132 \zeta_3)  \right. \nonumber \\
&& + 16 C_A^3 \left( 122043 T_r^2C_r^2 + 6776\frac{d^{abcd}_Fd^{abcd}_F}{d_A} \left( -11 + 24\zeta_3 \right)  \right) \nonumber \\
&& + 704 C_A^2 \left( 1521 T_r^2 C_r^3 + 112 T_r \frac{d^{abcd}_Fd^{abcd}_A}{d_A} \left( 4-39\zeta_3 \right) +242 C_r \frac{d^{abcd}_Fd^{abcd}_F}{d_A} \left( -11 +24 \zeta_3 \right)   \right) \nonumber \\
&& \left.  + 32 T_r C_A \left( 53361 T_r C_r^4 - 3872 C_r \frac{d^{abcd}_Fd^{abcd}_A}{d_A} \left( -4 +39\zeta_3 \right) + 112 T_r \frac{d^{abcd}_Ad^{abcd}_A}{d_A} \left( -5 +132 \zeta_3  \right) \right)  \right)
\end{eqnarray}
where $d_F^{abcd}$ and $d_A^{abcd}$ are a set of fully symmetrical tensors that can be found in \cite{vanRitbergen:1997va}. The dimension of the adjoint representation is $d_A$ while $\zeta_3$ is the Riemann zeta function evaluated at three. With these coefficients in hand we have at our disposal the exact scheme independent value of $\gamma_*$ through $O(\Delta_f^3)$.

\begin{figure}[t!]
  \centering
     \includegraphics[width=0.35\textwidth]{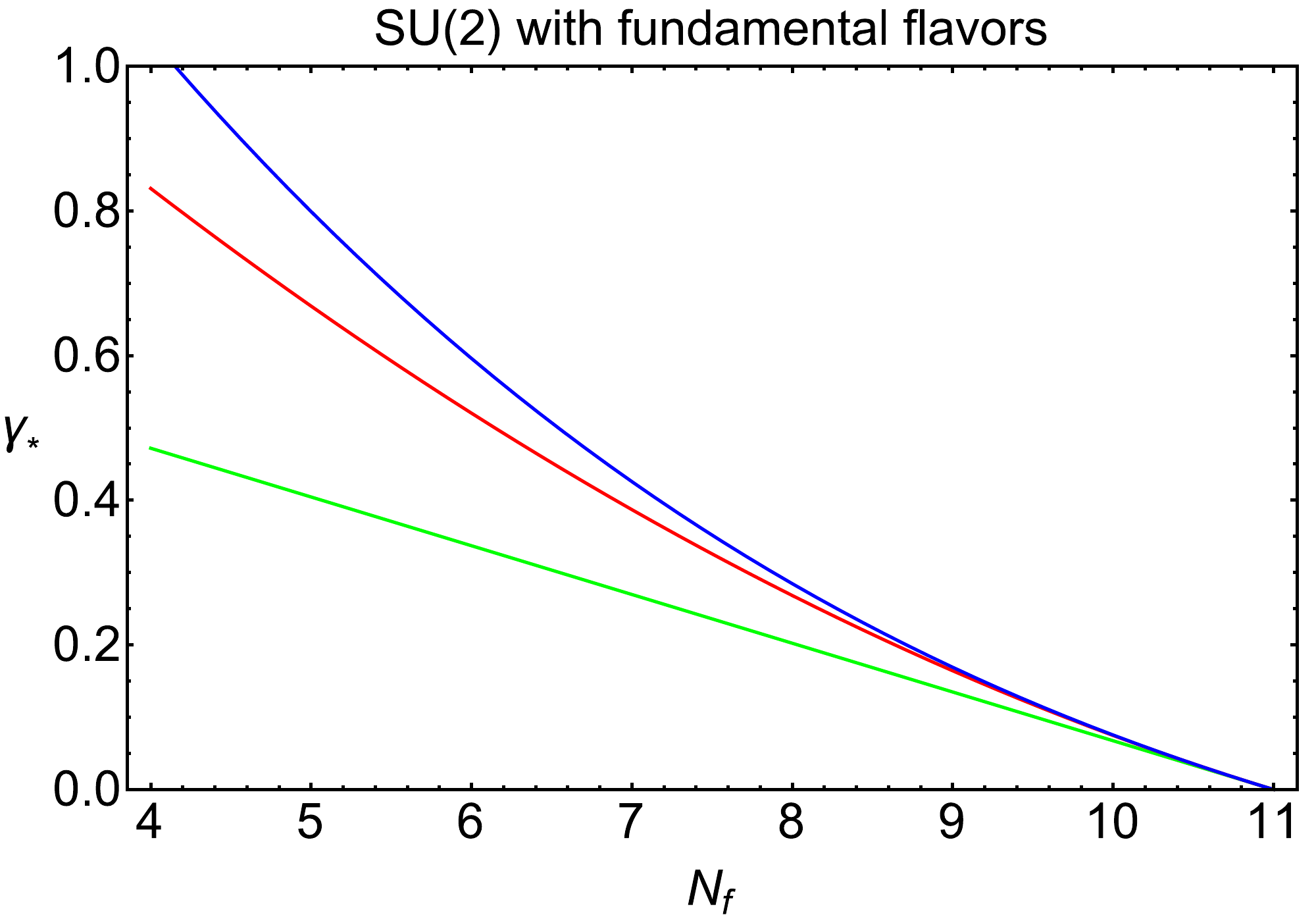}
     \qquad\qquad 
      \includegraphics[width=0.35\textwidth]{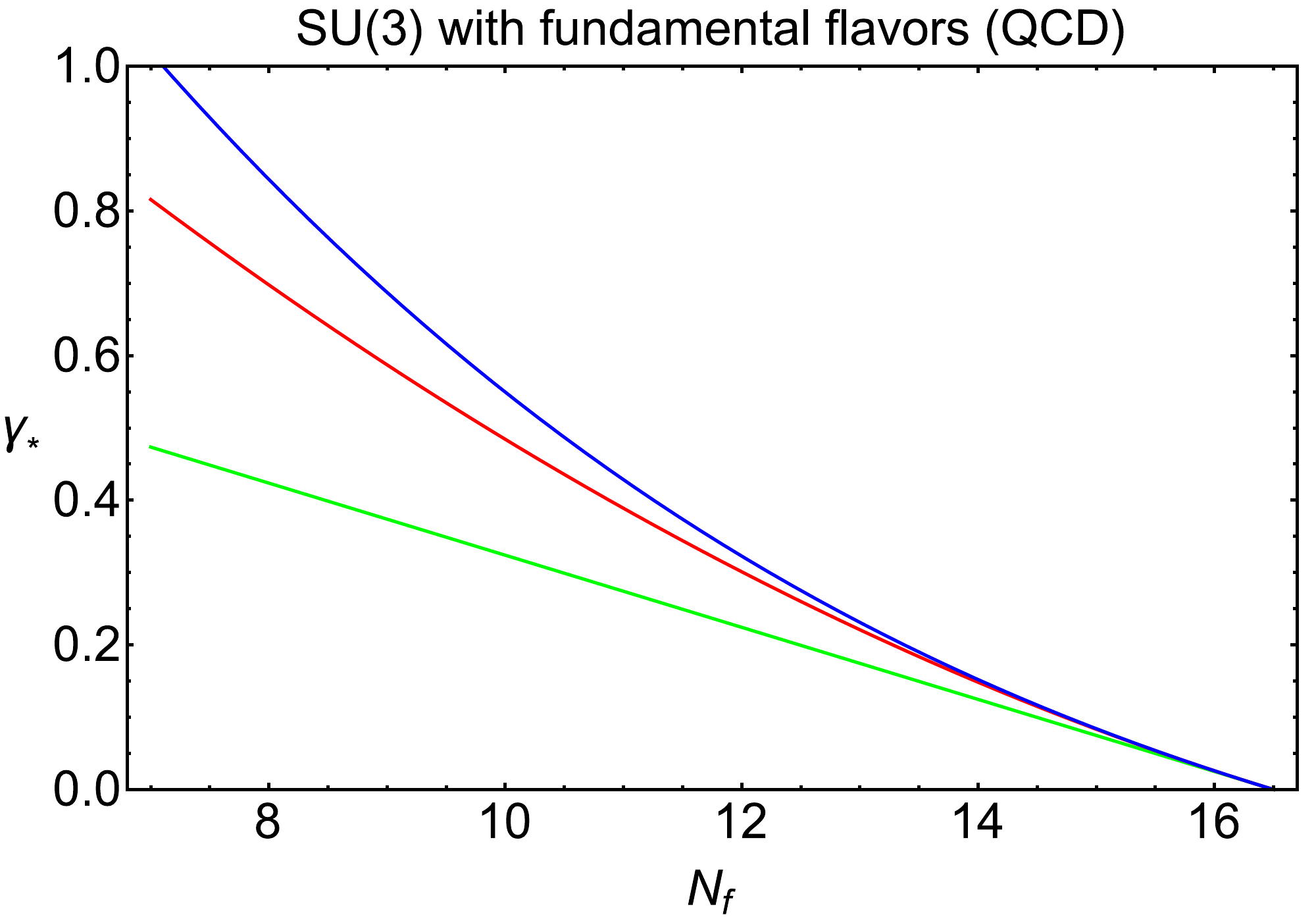}
   \caption{The value of $\gamma_*$ for a fermionic gauge theory with gauge group  $SU(2)$ (left) and $SU(3)$ (right) and  $N_f$ fundamental flavors. The green curve is $O(\Delta_f)$, the red curve is $O(\Delta_f^2)$ and the blue curve is $O(\Delta_f^3)$.}
   \label{F}
\end{figure}

\begin{figure}[t!]
  \centering
     \includegraphics[width=0.35\textwidth]{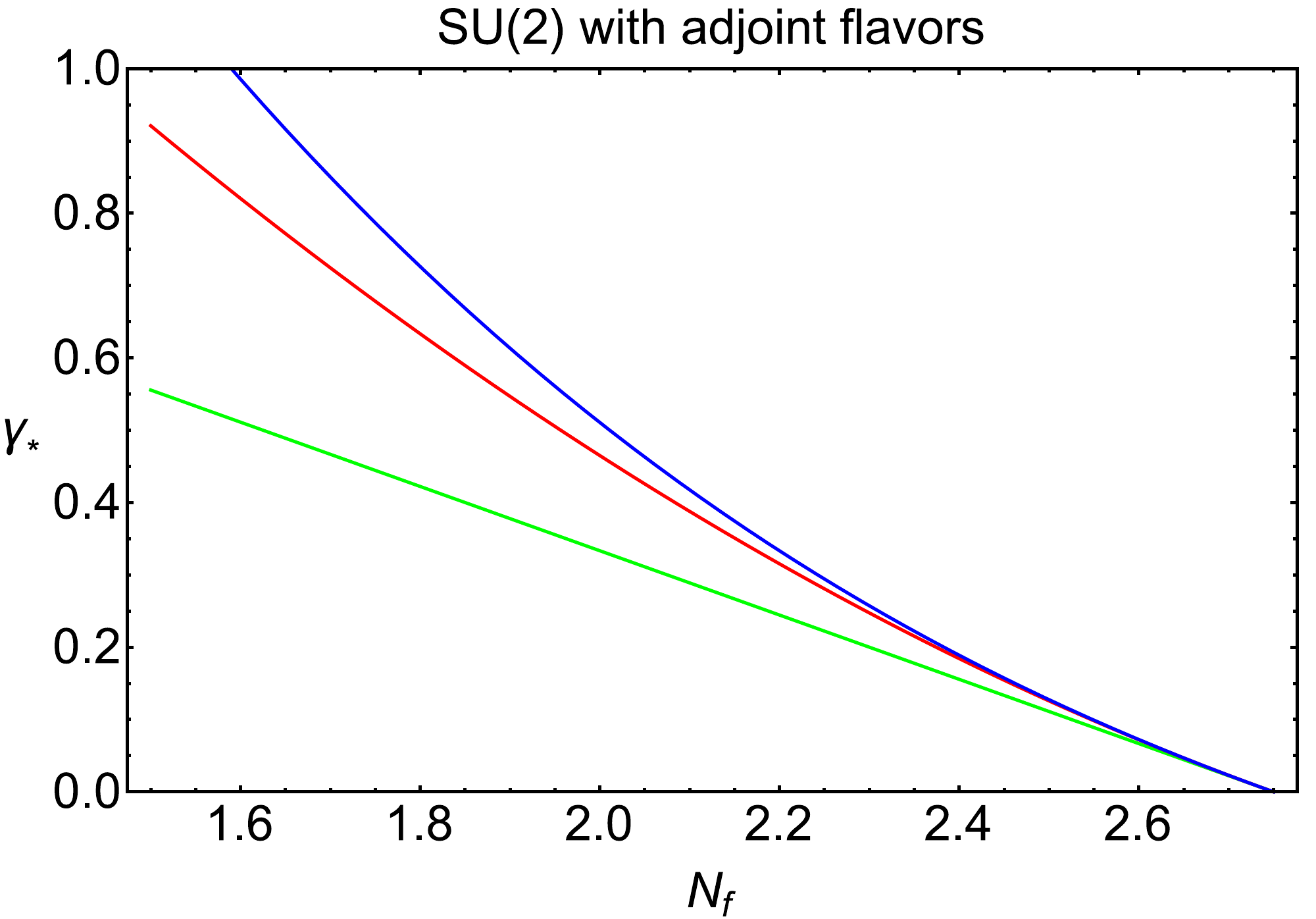}
     \qquad\qquad 
      \includegraphics[width=0.35\textwidth]{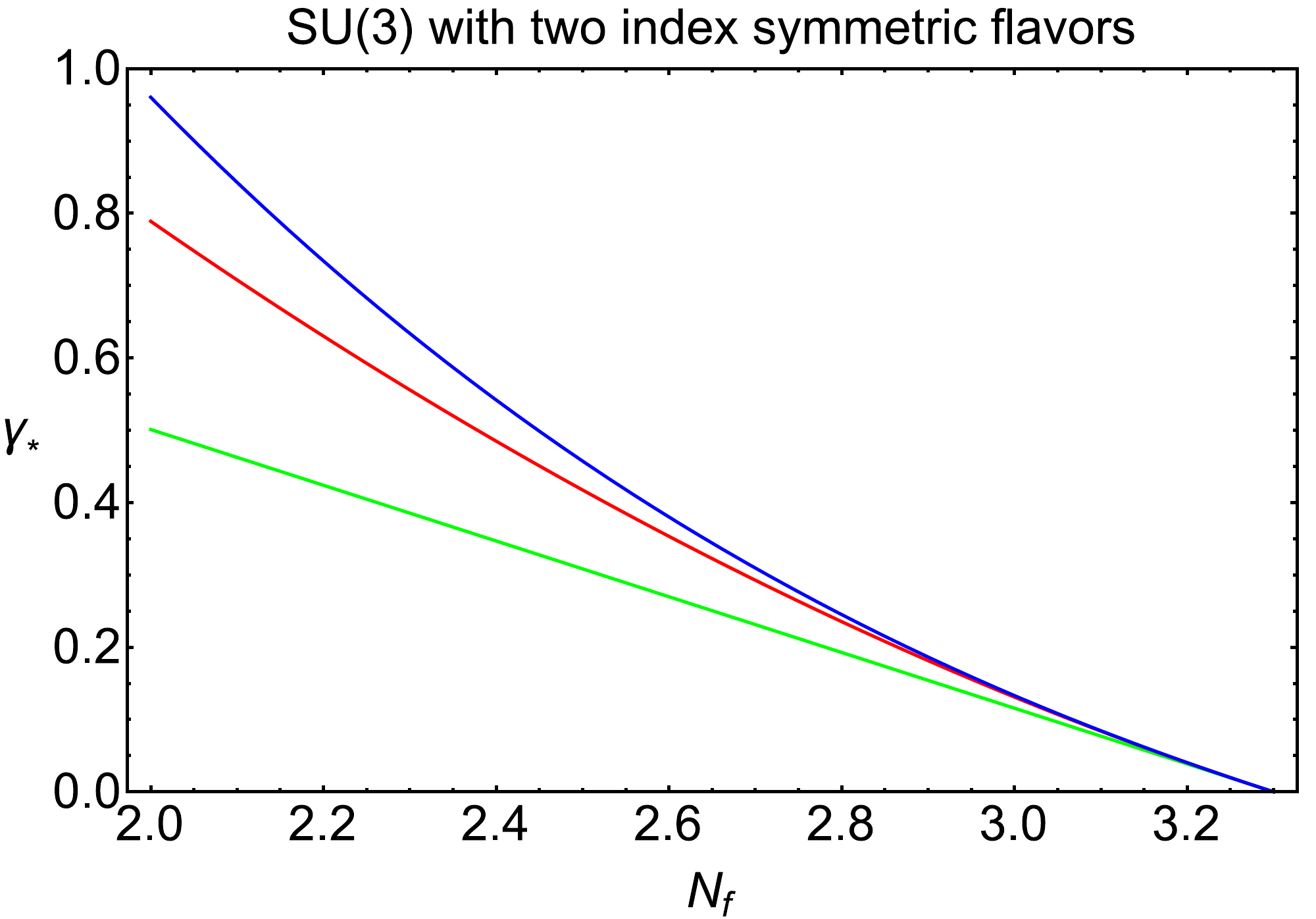}
   \caption{The value of $\gamma_*$ for a fermionic gauge theory with gauge group  $SU(2)$ (left) and $SU(3)$ (right) and  $N_f$ two index symmetric flavors. For two colors the two index symmetric representation is equivalent to the adjoint representation. The green curve is $O(\Delta_f)$, the red curve is $O(\Delta_f^2)$ and the blue curve is $O(\Delta_f^3)$.}
   \label{other}
\end{figure}

Again it is instructive to plot $\gamma_*$ as a function of the number of flavors. This we do for an $SU(2)$ and $SU(3)$ gauge group with fundamental fermions in Fig. \ref{F} and with two index symmetric fermions in Fig. \ref{other}. For fundamental flavors we observe that the value of  $\gamma_*$ stays relatively small for a large range of flavors before reaching unity.

In Tables \ref{tab:gamma-Fundamental} and \ref{tab:gamma-higher} we provide the value of $\gamma_*$ for a variety of theories. This includes $SU(3)$ QCD with fundamental fermions. If the conformal window is bounded by $\gamma_*<1$ then from eight flavors and up QCD is conformal. Although we expect that higher order corrections will push up the value of $\gamma_*$ slightly the conformal window extends quite far down in the number of flavors. 

There are two theories with higher dimensional representations that have received much attention as potential strongly interacting theories able to break the electroweak symmetry \cite{Sannino:2004qp}. The first is $SU(2)$ with two adjoint flavors. This theory has $\gamma_* \sim 0.511$ and must be assumed to lie within the conformal window. The second is $SU(3)$ with two two index symmetric flavors for which $\gamma_* \sim 0.960$. This theory seems to lie just around the boundary of the conformal window and could potentially exhibit walking dynamics.

\begin{table}[h]
\begin{center}
  \begin{tabular}{|c||c|c|c|c|c|c|c|}
\hline
  \multicolumn{8}{|c|}{{ $SU(2)$ }}  \\
 \hline\hline
 $N_f$ & 4 & 5 & 6 & 7 & 8 & 9 & 10  \\
  \hline
  $\gamma_*$ & 1.04 & 0.799 & 0.596 & 0.426 & 0.285 & 0.169 & 0.0754   \\ 
    \hline\hline
  \end{tabular}
  \\
\vspace{0.5cm}
 \begin{tabular}{|c||c|c|c|c|c|c|c|c|c|c|}
 \hline
  \multicolumn{11}{|c|}{{ $SU(3)$ }}  \\
 \hline\hline
 $N_f$ & 7 & 8 & 9 & 10 &11 & 12 & 13 & 14 & 15 & 16  \\
  \hline
  $\gamma_*$ & 1.02 & 0.844 & 0.687 & 0.549 & 0.428 & 0.323 & 0.231 & 0.152 & 0.0841 & 0.0259   \\ 
    \hline\hline
  \end{tabular}
  \caption{Values of $\gamma_*$ calculated through $O(\Delta_f^3)$ for $N_f$ fermions in the fundamental representation of an $SU(2)$ or $SU(3)$ gauge group. }
  \label{tab:gamma-Fundamental}
\end{center}
\end{table}

\begin{table}[h]
\begin{center}
  \begin{tabular}{|c||c|c|}
 \hline
  \multicolumn{3}{|c|}{{ $SU(2)$ }}  \\
 \hline\hline
 $N_f$ & 2  & $\frac{5}{2}$  \\
  \hline
  $\gamma_*$  & 0.511 & 0.127  \\ 
    \hline\hline
  \end{tabular}
\qquad \qquad 
 \begin{tabular}{|c||c|c|}
 \hline
  \multicolumn{3}{|c|}{{ $SU(3)$ }}  \\
 \hline\hline
 $N_f$ & 2 & 3  \\
  \hline
  $\gamma_*$  & 0.960 & 0.133  \\ 
    \hline\hline
  \end{tabular}
  \caption{Values of $\gamma_*$ calculated through $O(\Delta_f^3)$ for $N_f$ fermions in the two index symmetric representation of an $SU(2)$ or $SU(3)$ gauge group. For $SU(2)$ the two index symmetric representation is equivalent to the adjoint representation. }
  \label{tab:gamma-higher}
\end{center}
\end{table}

We also would like to make a comment on the convergence and accuracy of our result. In the supersymmetric case the ratio of two consecutive expansion coefficients is $\frac{c_{n+1}}{c_n} = \frac{2T_r}{4C_A}$. For supersymmetric QCD with $SU(3)$ gauge group and fundamental superflavors this is $\frac{c_{n+1}}{c_n}  \sim 0.11$. In the non-supersymmetric case on the other hand we find $\frac{c_2}{c_1} \sim 0.076$ and $\frac{c_3}{c_2} \sim 0.063$. Hence the first expansion coefficients decrease more rapidly in the non-supersymmetric case compared to the supersymmetric case. If this continues to higher orders the radius of convergence and the accuracy of our results will be even more favourable than in the supersymmetric case. The same pattern holds for the other gauge groups and representations discussed here. 

We stress that the above theories have in recent years been subject to thorough studies by the lattice community. For a recent review see \cite{DeGrand:2015zxa}. Our investigations should serve as an analytic background against which the lattice simulations should be compared. Having in mind the high level of accuracy we have seen to exist for supersymmetric theories we cannot help but speculate that the computation of $\gamma_*$ through $O(\Delta_f^3)$ is at least equally precise for QCD and similar non-supersymmetric fermionic gauge theories.

Finally we make a brief comment on the adjoint theory. Occasionally it has been speculated that the physics of an $SU(N)$ gauge theory with a set of adjoint flavors inside the conformal window should be independent of $N$. We can finally show that this is not the case. Although $c_1 =\frac{4}{9}$ and $c_2= \frac{341}{1458}$ both do not depend on $N$ actually $c_3 = \frac{61873-\frac{42624}{N^2}}{472392}$ has a mild dependence on $N$. Note that $\zeta_3$ drops out. Since these coefficients are exact $\gamma_*$ is bound to have at least some minor $N$ dependence. Potentially this could change the boundary of the conformal window as a function of $N$.

\section{Conclusion}\label{sec:conclusion}

We have proposed a novel way to consistently calculate the anomalous dimension $\gamma_*$ at a fixed point as a series expansion in $\Delta_f$ to any finite order in perturbation theory. Done in this way any scheme dependence that would show up using the standard approach is eliminated. Using the $n+1$ loop beta function and $n$ loop anomalous dimension allow us to calculate $\gamma_*$ in an exact and scheme independent manner through $O(\Delta_f^n)$. We then crosschecked with exact results in supersymmetric QCD and found agreement. We observed that already at $O(\Delta_f^3)$ the perturbative calculation of $\gamma_*$ provides a surprisingly accurate result. Our computation preserves supersymmetry order by order. Finally we computed $\gamma_*$ for QCD at $O(\Delta_f^3)$ using the available four loop beta function and three loop anomalous dimension. Small values of $\gamma_*$ is the hallmark of these theories for a large range of flavors.

{\it Acknowledgments:}  The CP$^3$-Origins center is partially funded by the Danish National Research Foundation, grant number DNRF90. The author would like to thank C. Pica, F. Sannino and R. Shrock for valuable discussions and comments.

\end{document}